\documentclass[article]{jss} 

\usepackage{orcidlink,thumbpdf,lmodern}
\usepackage{framed}
\usepackage{amsmath,amsthm,amssymb}
\usepackage{mathabx}
\usepackage{tabularx}
\usepackage{latexsym}
\usepackage{accents} 
\usepackage{amsfonts}
\usepackage{mathtools} 
\usepackage{bm}
\usepackage{dsfont}
\usepackage{bbm}
\usepackage{sgame}
\usepackage[shortlabels]{enumitem}
\usepackage[para,online,flushleft]{threeparttable}	
\usepackage[linesnumbered,lined,ruled,commentsnumbered]{algorithm2e}

\def\vector#1{\mbox{\boldmath{$#1$}}} 
\def\matrix#1{\mathbf #1} 
\newcommand{\inv}{^{-1}}
\newcommand{\one}{\mathds{1}}

\author{Annalivia Polselli~\orcidlink{0009-0002-7579-7926}\\Institute for Social and Economic Research, University of Essex}
\Plainauthor{Annalivia Polselli}

\title{xtdml: Double Machine Learning Estimation to Static Panel Data Models with Fixed Effects in \proglang{R}}
\Plaintitle{xtdml: Extending Double Machine Learning Estimation to Static Panel Data Models with Fixed Effects in R}
\Shorttitle{Double Machine Learning Estimation for Static Panel Data Models in R}

\Abstract{The double machine learning (DML) method combines the predictive power of machine learning with statistical estimation to conduct inference about the structural parameter of interest. This paper presents the \proglang{R} package \pkg{xtdml}, which implements DML methods for partially linear panel regression models with low-dimensional fixed effects, high-dimensional confounding variables, proposed by \citet{clarke2025}. The package provides functionalities to: (a) learn nuisance functions with machine learning algorithms from the \pkg{mlr3} ecosystem, (b) handle unobserved individual heterogeneity choosing among first-difference transformation, within-group transformation, and correlated random effects, (c)  transform the covariates with min-max normalization and polynomial expansion to improve learning performance. We showcase the use of \pkg{xtdml} with both simulated and real longitudinal data.}


\Address{%
  Annalivia Polselli\\
  Institute for Social and Economic Research\\
  University of Essex\\
  Wivenhoe Park, CO4 3SQ\\
  United Kingdom\\
  E-mail: \email{annalivia.polselli@essex.ac.uk}\\
  URL: \url{https://sites.google.com/site/annaliviapolselli/home?authuser=0}
}

\begin{document}

\section[Introduction]{Introduction} \label{sec:intro}
Double machine learning (DML) is a  method for causal inference that leverages machine learning tools to flexibly model the functional form of the covariates while preserving valid statistical inference for the structural parameter (or treatment effect).
Several works have contributed to the development of the DML method across a wide range of settings (i.e., treatment endogeneity, treatment heterogeneity, binary and continuous treatments) in cross‐sectional models \citep{chernozhukov2018,lewis2020,knaus2022double,huber2023,scaillet2025,chernozhukov2025}, static and dynamic panel data models \citep{klosin2022,semenova2023,arganaraz2025,clarke2025,marquez2025}, and difference-in-differences (DID) designs \citep{chang2020,haddad2024,zhang2025}. The availability of \proglang{R} \citep{R} packages for some of these methods has further contributed to the dissemination and practical adoption of DML in applied research.\footnote{A growing number of empirical studies have applied these DML methods to estimate treatment effects, including, for example, \citet{deryugina2019}, \citet{langen2023}, \citet{strittmatter2023}, \citet{baiardi2024,baiardi2024plough}, and \citet{li2025}.}  

This article introduces the \proglang{R} package \pkg{xtdml} \citep{xtdml}, which implements the DML method for static panel data models with fixed effects as developed in \citet{clarke2025}. The package fits partially linear panel regression (PLPR) models with low-dimensional fixed effects and treatment exogeneity, and permits flexible adjustment for high-dimensional confounding. To account for unobserved individual heterogeneity, the user can choose among the following panel data approaches:

\begin{itemize}
    \item within-group approximation (WG-approximation) transformation, 
    \item first-difference exact (FD-exact) transformation,
    \item correlated random effects (CRE) device.
\end{itemize}

The \pkg{xtdml} package builds on the object-oriented architecture of \pkg{DoubleML} \citep{DoubleML}, which provides a comprehensive and user-friendly toolkit for the estimation of treatment effects in cross-sectional partially linear and interactive regression models within the DML framework.\footnote{The \pkg{xtdml} package can be viewed as the panel data counterpart of \pkg{DoubleML} for estimating partially linear regression models with DML, in the same way as \pkg{plm} \citep{plm} is the panel data analogue of the \code{lm()} function in the \pkg{stats} \citep{stats} package for linear regression. For more details on  \pkg{DoubleML} see \citep[][]{bach2024}.}  \pkg{xtdml} adopts the \pkg{R6} class system \citep{R6} and integrates with the \pkg{mlr3} \citep{mlr3} ecosystem to support state-of-the-art learners and tuning methods.
The core model class \code{xtdml_plr} supports estimation within PLPR models with fixed effects by learning nuisance components through machine learning and constructing Neyman-orthogonal score functions based on adapted moment conditions. 
Common tasks implemented by \pkg{xtdml} include the estimation of the target parameter, the calculation of standard errors, t-tests, confidence intervals, and root mean squared errors (RMSEs). Similarly to \pkg{DoubleML}, the package offers users control over the estimation procedure, including  the choice of learners for each nuisance function, hyperparameter tuning, the specification of the DML algorithm, and the selection of orthogonal score functions appropriate to the panel data setting. 
In contrast to the original \pkg{DoubleML}, \pkg{xtdml} introduces several features tailored to panel data including:
\begin{enumerate}
    \item \textbf{Panel data approaches.} Users can choose among several panel data approaches (within-group, first-differences, correlated random effects) which transform both the data frame and the associated learning tasks accordingly. Alternatively, the user may select the option for a pooled estimation approach, which implements the original DML by \citet{chernozhukov2018} with cluster-robust inference and no panel data transformation is applied to the structural model; this approach is suitable for repeated cross-section data. 
    \item \textbf{Covariate pre-processing.} Users can select the transformation of the included covariates, with options including polynomial expansion, min-max normalization, or no transformation.\footnote{Each transformation is suitable for specific base learners to improve learning performance. A standard approach is to construct an extensive dictionary of the covariates when running Lasso regressions (a polynomial expansion is thus preferable) while normalising the scale of the covariates (inputs) is common with Neural Networks through min-max transformation or standardisation. Finally, no transformation of the variables is required for tree-based learners, which can capture interaction terms.}
    \item \textbf{Panel structure specification.} Users are required to specify  panel and time identifiers to correctly apply one of the selected panel data transformations and obtain statistical inference. If the user requires cluster‐robust standard errors at a level different from the panel identifier, an alternative clustering variable can be specified. 
    \item \textbf{Sample splitting with cross-fitting for panel data.} The implementation supports \emph{block-k-fold} sample splitting, where each subject with its entire time series is assigned to a unique fold. This ensures that resampling is performed at the subject level rather than at the observation level, as preferable with panel data. 
    \item \textbf{Model evaluation.} The package provides measures of predictive performance of the base learners, such as the root mean squared error (RMSE) for the overall model and for the different nuisance parameters.
\end{enumerate}

\pkg{xtdml} is part of a limited class of \proglang{R} packages that allow the estimation of longitudinal or panel data models with machine learning techniques. Other \proglang{R} packages include the following: \pkg{ablasso} \citep{ablasso} combines the Arellano-Bond estimation method with the Least Absolute Shrinkage and Selection Operator (Lasso) for the estimation of structural parameters in dynamic linear panel models with lagged dependent variables and predetermined covariates as explanatory variables, and unobserved individual and time fixed effects \citep{chernozhukov2024}. Among the class of causal estimators implemented in \pkg{causalweight} \citep{causalweight},  \code{didcontDML} and \code{didcontDMLpanel} functions conduct the estimation of the average treatment effect on the treated using the DID methods for repeated cross-sections and panel data, respectively, while flexibly controlling for confounding via DML \citep[based on][]{chang2020,haddad2024}.\footnote{Similar to \pkg{causalweight}, \pkg{xtdml} can accommodate a classical DID design with two time periods and two groups (or a two-way fixed-effects setup) without staggered adoption by specifying \code{approach = "fd-exact"} in \code{xtdml\_data\_from\_data\_frame()} and using $Treat_i \times Post_t$ as the treatment variable. The advantage of \pkg{xtdml} over \pkg{causalweight} is the direct integration with the \pkg{mlr3} ecosystem, which provides access to a broad range of base learners.}
The package \pkg{desla} \citep{desla} supports the implementation of desparsified Lasso with both cross-sectional data \citep{van2014} and high-dimensional time series \citep[based on][]{adamek2023,adamek2024}.
\pkg{REEMtrees} \citep{REEMtree} fits a single regression tree with individual-specific effects \citep[based on][]{sela2012}, designed for predict models with longitudinal or clustered data. 

The \pkg{hdm} package \citep{hdm} implements double-selection procedures for high-dimensional cross-sectional settings, fitting Lasso regressions with non-Gaussian errors, with or without instrumental variables. Although originally designed for cross-sectional data, it can be adapted to panel data settings by manually transforming the data through conventional panel data transformations, such as within-group or first-differences. This is possible because the method relies on penalized and linear regressions, provided that the user specifies an \emph{extensive dictionary} of covariates (such as high-order polynomials and interaction terms) to consistently estimate the parameter of interest. In contrast, the \pkg{DoubleML} toolkit for cross-sectional models cannot be used directly with panel data settings in the same way as \pkg{hdm} as it allows for nonlinear machine-learning learners, for which standard panel data transformations are not valid without further methodological and software adjustments, which are now implemented in \pkg{xtdml}. 

\begin{table}[t!]
\centering
\scalebox{.65}{
\begin{threeparttable}	
   \begin{tabular}{llcl}  
    \hline
    R package&Description&Causal analysis&Base learners\\	
    \hline
    \vspace{-2mm}\\
     \pkg{ablasso}& Combines Arellano-Bond estimator with Lasso&Yes& Lasso\\
     \pkg{causalweight} &Implements DID within DML framework&Yes& Lasso, SVM, CART, ensemble\\
      \pkg{desla}& Implements desparsified (post-)Lasso procedure&Yes& Lasso\\
     \pkg{hdm}&Implements double selection procedures&Yes&(Post-)Lasso\\      
     \pkg{REEMtree}&Incorporates mixed effects models in tree-based estimation methods &No &modified CART\\
     \pkg{xtdml}&Fits partially linear panel regression models within DML &Yes&
     any from \pkg{mlr3}\\
\hline
\end{tabular} 
Note: CART stands for classification and regression tree; SVM stands for support vector machine. \pkg{hdm} is not originally developed for panel data, but it can be accommodated for such data by manually transforming the data using standard panel transformations, such as within-group or first-difference transformations.
\end{threeparttable}
}
\caption{Summary of \proglang{R} packages for machine learning and longitudinal / panel data.}\label{tab:rpack}
\end{table}

The rest of the article is structured as follows. Section~\ref{sec:models} introduces the partially linear regression model with fixed effects and its transformed versions under three panel data approaches. Section~\ref{sec:dml} presents the DML algorithm for panel data, and discusses its main elements  by panel data approach: the learning procedure, the construction of Neyman orthogonal score functions, and \emph{block-k-fold} sample splitting. Section~\ref{sec:learners} discusses hyperparameter tuning and the transformation of the variables allowed in the package. Section~\ref{sec:examples} provides a guide of the main functions and methods in the package with simulated data. Section~\ref{sec:empirical} illustrates the use of the package with real data. Last section concludes.


\section{Partially Linear Regression Model with Fixed Effects} \label{sec:models}
 Consider the partially linear panel regression (PLPR) model with fixed effect $\alpha^*_i$ 
\begin{equation}\label{eqn:plr-iv}
Y_{it} = D_{it}\theta_0  + g_1(\vector X_{it}) + \alpha^*_i + U_{it},
\end{equation}
where $g_1$ is a non-linear nuisance function of $\vector X_{it}$ and $\mathbb{E}[U_{it}|D_{it},\vector X_{it},\alpha^*_i]=0$ and $\mathbb{E}[\alpha_i^*\mid D_{it},\vector X_{it}]\neq 0$.  The target parameter $\theta_0$ is the average partial effect (APE) of continuous $D_{it}$ such that $d\theta_0=\mathbb{E}[Y_{it}(d)-Y_{it}(0)]$, or the average treatment effect (ATE) $\mathbb{E}[Y_{it}(1)-Y_{it}(0)]$ for binary treatments.

The partial-out PLPR (PO-PLPR) model is
\begin{align}
Y_{it} &= V_{it}\theta_0  + l_1(\vector X_{it}) + \alpha_i +  U_{it}\label{eqn:plr_y},\\
V_{it}&= D_{it} - m_1(\vector X_{it}) -\gamma_i,\label{eqn:plr_v}
\end{align}
where $l_1$ and $m_1$ are nuisance functions, $\alpha_i$ is a fixed effect, $\mathbb{E}[U_{it}\mid V_{it},\vector X_{it},\alpha_i]=0$, and $V_{it}$ the residual of a non-linear additive noise treatment model depending on fixed effect $\gamma_i$ and satisfying $\mathbb{E}[V_{it}|\vector X_{it},\gamma_i]=0$. 

Models~\eqref{eqn:plr-iv}–\eqref{eqn:plr_v} cannot be directly estimated from the observed data due to the presence of unobserved individual heterogeneity, which is assumed to be correlated with the covariates and the treatment. Ignoring the panel structure and proceeding with a pooled regression without accounting for the fixed effects would make the double machine learning estimator inconsistent due to omitted variable bias. As standard practice in Econometrics, unobserved individual heterogeneity can either be eliminated through panel data transformations (through within-group or first-differences transformation)  or explicitly modelled (correlated random effects).

In the following sections, we discuss the panel data approaches implemented in the \pkg{xtdml} package to address the presence of the unobserved individual heterogeneity. These are based on the panel data  approaches adapted for DML by \citet{clarke2025}: correlated random effects (CRE),  first-differences exact (FD exact) transformation, and  within-group approximation (WG approximation) transformation. 

\subsection{The CRE Approach (Mundlak's Device)}
The first approach models unobserved individual heterogeneity through the Mundlak device \citep{mundlak1978}, converting the fixed-effects model into a random-effects specification. Under additive separability
\begin{enumerate}[label=(\alph*)]
    \item  $\mathbb{E}[Y_{it}(0)\mid\vector X_{it},\overline{X}_i]=g_1(\vector X_{it})+\alpha^*_i$   where $\alpha^*_i=\alpha^*(\overline{X}_i),$  \label{item:plpr}
    \item  $\mathbb{E}[D_{it}\mid\vector X_{it},\overline{X}_i]=m_1(\vector X_{it})+\gamma_i$        where $\gamma_i=\gamma(\overline{X}_i)$,\label{item:po-plpr}
\end{enumerate}
\noindent  $\mathbb{E}[Y_{it}(0)\mid \vector X_{it}, \overline{X}_i]= \mathbb{E}[Y_{it}\mid\vector X_{it},\overline{X}_i]- \mathbb{E}[D_{it}\mid\vector X_{it},\overline{X}_i]\theta_0$. 

The PO-PLPR model follows because $\mathbb{E}[Y_{it}\mid\vector X_{it},\overline{X}_i]=l_1(\vector X_{it})+\alpha_i$, where $l_1(\vector X_{it})=g_1(\vector X_{it})+m_1(\vector X_{it})\theta_0$ and $\alpha_i=\alpha^*_i+\gamma_i\theta_0$. Then, the PO-PLPR model for CRE with nuisance parameters $\widetilde{l}_1$ and $\widetilde{m}_1$ is
\begin{align}
    Y_{it} & =V_{it}\theta_0 + \widetilde{l}_1(\vector X_{it},{\overline {\vector X}}_i)+a_i+U_{it} \label{eqn:cre_y}\\ 
    V_{it} & = D_{it} - \widetilde{m}_1(\vector X_{it},{\overline{\vector X}_i}) - c_i, \label{eqn:cre_v} 
\end{align} 
\noindent where $\mathbb{E}[U_{it} \mid V_{it},\vector X_{it},{\overline{\vector X}_i},a_i\big]=\mathbb{E}[V_{it}|\vector X_{it},{\overline{\vector X}_i},c_i]=0$, $\widetilde{l}_1(\vector X_{it},{\overline{\vector X}_i})=l_1(\vector X_{it}) + \omega_{\alpha}({\overline{\vector X}_i})$ and $\widetilde{m}_1(\vector X_{it},{\overline{\vector X}_i}) = m_1(\vector X_{it})+\omega_\gamma({\overline{\vector X}_i})$. The random effects, $a_i$ and $c_i$, capture autocorrelation between observations on the same individual. The nuisance parameters $\widetilde{l}_1$ and $\widetilde{m}_1$ can be learnt directly from $\mathbb{E}[Y_{it}\mid\vector X_{it}, {\overline{\vector X}_i}]$ and $\mathbb{E}[D_{it}\mid\vector X_{it}, {\overline{\vector X}_i}]$, respectively.

Similarly, the PLPR model for CRE is
\begin{equation}\label{eqn:cre_iv} 
Y_{it} = D_{it}\theta_0  + \widetilde{g}_1(\vector X_{it},{\overline{\vector X}_i}) +a_i+U_{it},
\end{equation}
where the nuisance parameter $\widetilde{g}_1$ can be learnt iteratively from $Y_{it} - \mathbb{E}[Y_{it}\mid\vector X_{it}, {\overline{\vector X}_i}]$.

\subsection{The Transformation Approaches (FD and WG)}
The second type of approaches follows more conventional estimation techniques for panel data, where the unobserved individual heterogeneity is removed from the model by transforming the data. Let $Q$ be a panel data transformation operator, the within-group (WG) or time-demeaning transformation is $Q(W_{it})=W_{it}-\overline{W}_i$ for a generic random variable $W_{it}$, where $\overline{W}_{i}= T\inv \sum_{t=1}^T W_{it}$. The first-difference (FD) transformation is $Q(W_{it}) = W_{it}-W_{it-1}$ for $t=2,\dots,T$. 

The transformed models for $Q(Y_{it})$ and $Q(V_{it})$ under PO-PLPR model~\eqref{eqn:plr_y}-\eqref{eqn:plr_v} is 
\begin{align}
    Q(Y_{it}) & = Q(V_{it})\theta_0 + Q\big({l}_1(\vector X_{it})\big) + Q(U_{it}) \label{eqn:Q_y} \\
    Q(V_{it}) & = Q(D_{it})-Q\big({m}_1(\vector X_{it})\big), \label{eqn:Q_d}
\end{align}
and under PLPR model~\eqref{eqn:plr-iv} is
\begin{equation} \label{eqn:Q_iv}
Q(Y_{it}) = Q(D_{it})\theta_0  + Q\big(g_1(\vector X_{it})\big) + Q(U_{it}),
\end{equation}
\noindent which do not depend on fixed effects $\alpha_i$ and $\gamma_i$ because $Q(\alpha_i)=Q(\alpha^*_i)=Q(\gamma_i)=0$. The transformed nuisance functions $\{Q\big(l_1(\vector X_{it})\big),Q\big(m_1(\vector X_{it})\big)\}$  can be learnt directly from the data while $Q\big(g_1(\vector X_{it})\big)$ can be learnt iteratively from $Q(Y_{it}) - \mathbb{E}[Q(Y_{it})\mid Q(\vector X_{it})]$. 

In the following section, we discuss how the learning procedure is implemented in \pkg{xtdml} from available data.

\section{Double Machine Learning for Panel Data Models}\label{sec:dml}
The objective is to make inferences on the target parameter $\theta_0$, given suitable predictions of the vector of nuisance parameters $\bm{\eta}_0=(l_0,m_0,g_0)$ obtained using algorithms from the \pkg{mlr3} ecosystem (e.g., \pkg{mlr3}, \pkg{mlr3learners}, \pkg{mlr3extralearners}).
The estimation procedure implemented in \pkg{xtdml} accounts for (a) the presence of the unobserved individual heterogeneity correlated with the observed factors, and (b) possibly nonlinear functions of the covariates to obtain a consistent estimate of the structural (causal) parameter. 

The DML procedure presented in this section relies on three key elements. First, the nuisance functions are predicted (or \emph{learnt}) accounting for both (a) and (b). Second, Neyman orthogonal score functions are constructed based on moment conditions derived from models presented in Section~\ref{sec:models} to reduce the first-order bias introduced by regularized learners (low variance, but large bias). Third, the sample is divided into $K$ groups (or folds), where a part is used for predictions and the rest for estimating the target parameter to reduce overfitting bias (low bias, but large variance). The employed splitting scheme respects the longitudinal structure of the data by randomly assigning a subject with its entire time series to a unique fold $k$ (block-k-fold sample splitting).

Algorithm~\ref{alg:xtdml} below summarizes the DML procedure implemented in \pkg{xtdml}, which adapts Algorithms~1 and~2 in \citet{chernozhukov2018} and \citep{bach2024} to longitudinal datasets. The argument \code{dml_procedure} in the \code{new()} method specifies the DML algorithm to implement (\code{dml_procedure = c('dml1', 'dml2')}, with \code{dml2} as the default). 

\begin{algorithm}[ht!]
\SetAlgoLined
\SetKwData{Left}{left}
\SetKwData{This}{this}
\SetKwData{Up}{up}
\SetKwFunction{Union}{Union}
\SetKwFunction{FindCompress}{FindCompress}
\SetKwInOut{Input}{Input}
\SetKwInOut{Output}{Output}
\SetKwInOut{Require}{Require}
\SetKwInOut{State}{State}
\SetKwInOut{Initialize}{Initialize}

\caption{Panel DML Algorithm}\label{alg:xtdml}
\vspace{2mm}
\Require{Dataset with $N$ cross-sectional units, and $T\ge 2$ time periods. }
\Input{Data $\{Y_{it},D_{it},\vector X_{it}\}$, panel and time identifiers; cluster variable if different from panel identifier.}
\Output{$\widehat{\theta}$, $\widehat{\Sigma}$, model RMSE, RMSE~of~$l$,~$m$ and $g$ if \texttt{score = 'IV-type'}.}
\Initialize{Set number of folds $k\ge2$. Select the type of score function \texttt{score = {c('PO-type', 'IV-type')}}, type of panel data approximation \texttt{approx = {c('fd-exact', 'wg-approx', 'cre', 'pooled')}}, and the type of transformation of $X$ \texttt{transformX = {c('no', 'poly', 'minmax')}}, and the double machine learning procedure \code{dml\_procedure = c('dml1', 'dml2')}. Assign base learners to each nuisance parameters $\bm{\eta}  =  \{l, m, g\}$ from \pkg{mlr3} ecosystem and define tuning settings.}

\vspace{3mm}
Data is sorted by \code{panel_id} and \code{time_id}, transformed according to the choice of \code{approx} and \code{transformX}.

The sample is divided into $K$ folds, where subject $i$ is randomly assigned to fold $k$ $W_k$ is the estimation sample for fold $k$ and  of size $|N_k|$; $W_k^c$  is the complementary sample for folds $-k$ and of size $|N-N_k|$.

\eIf{\texttt{dml\_procedure  =  "dml1"}}
{
     \For{$k\leftarrow 1$ \KwTo $K$}{
        Predict $\widehat{\bm{\eta}}_k$ using base learners on data $\mathcal{W}_k^c$ .\\
        Construct Neyman orthogonal score function $\mathbf{s}(W_k, \theta_k; \widehat{\bm{\eta}}_k)$.\\
        Solve $\frac{1}{|N_k|}\sum_{i\in \mathcal{W}_k}\mathbf{s}(W_k, \theta_k; \widehat{\bm{\eta}}_k) = 0$ for $\theta_k$ using data $W_k$.
    }
    Average estimate of $\{\widehat{\theta}_k\}_{k = 1}^K$ and model RMSE, compute finite-sample variance-covariance matrix $\widehat{\bm{\Sigma}}$, and  RMSE~of~$\{l,r,g\}$. 
}{
    \If{\texttt{dml\_procedure  =  "dml2"}}
    {
        \For{$k\leftarrow 1$ \KwTo $K$}{
        Predict $\widehat{\bm{\eta}}_k$ using base learners on data $\mathcal{W}_k^c$ .\\
        Construct Neyman orthogonal score function $\mathbf{s}(W_k, \theta_k; \widehat{\bm{\eta}}_k)$.\\
        Solve $\frac{1}{|N_k|}\sum_{k = 1}^K\sum_{i\in \mathcal{W}_k}\mathbf{s}(W_k, \theta_k; \widehat{\bm{\eta}}_k) = 0$ for $\theta_k$ using data $W_k$.
        }
        Final estimate of $\{\widehat{\theta}_k\}_{k = 1}^K$ and model RMSE, compute finite-sample variance-covariance matrix $\widehat{\bm{\Sigma}}$, and  RMSE~of~$\{l,r,g\}$.
    }
}

\end{algorithm}

\subsection{Learning Nuisance Functions}\label{sec:learning}
The first stage of the DML procedure consists of learning the nuisance functions, $\bm{\eta}_0=(l_0,m_0,g_0)$,  using machine learning algorithms while accounting for adjustment due to the longitudinal structure of the data. 
The  \code{approach}  argument in the \code{xtdml_data_from_data_frame} function allows the user to choose among \code{("fd-exact", "wg-approx", "cre", "pooled")}, where no default option is provided.\footnote{As previously mentioned, the pooled estimation approach implements the cross-sectional DML estimation with cluster-robust statistical inference, as in \pkg{DoubleML}. This option is recommended for repeated cross-sections.} The data is transformed according to the selected approach using the specified panel and time identifiers, and the prediction tasks for the nuisance parameters are adapted accordingly. 

The prediction tasks (or \emph{learning stage}) implemented in the \pkg{xtdml} package differ by panel data approach.

\textbf{The CRE Approach (Mundlak Device).} The procedure is implemented by selecting the argument \code{approx = "cre"} in the function \code{xtdml_data_from_data_frame}, which internally generates the individual means of the covariates by \code{panel_id}. The nuisance function
    \begin{itemize}
    \item $\widetilde{l}_1(\vector X_{it},\overline{\vector X}_i)+a_i$ is learnt directly from transformed data $\{Y_{it},\vector X_{it},\overline{\vector X}_i:t=1,\dots,T\}_{i=1}^N$,
    \item $\widetilde{m}_1(\vector X_{it},\overline{\vector X}_i)+c_i$ is learnt directly from transformed data $\{D_{it},\vector X_{it},\overline{\vector X}_i,\overline{D}_i:t=1,\dots,T\}_{i=1}^N$  to construct the orthogonal estimator $ V_{it}= D_{it} - \widetilde{m}_1(\vector X_{it},\overline{\vector X}_i) - c_i$,
    \item $\widetilde{g}_1(\vector X_{it},\overline{\vector X}_i)+a_i$ is learnt iteratively from $\{Y_{it},\vector X_{it},\overline{\vector X}_i:t=1,\dots,T\}_{i=1}^N$ for model~\eqref{eqn:plr-iv}, relying on the estimates of $\widetilde{l}_1$ and $\widetilde{m}_1$.
\end{itemize}

\textbf{The WG Approximation Approach.} The procedure is implemented by selecting the argument \code{approx = "wd-approx"}, which internally time-demeans all variables by \code{panel_id}. The nuisance function
\begin{itemize}
    \item $Q\big({l_0(\vector{X}_{it})}\big)\approx {l}_1\big({Q(\vector{X}_{it})}\big)$ is learnt directly from transformed data $\{Q(Y_{it}), Q(\vector{X}_{it})\}$,
    \item $Q\big({m_0(\vector{X}_{it})}\big)\approx {m}_1\big({Q(\vector{X}_{it})}\big)$ is learnt directly from transformed data $\{Q(D_{it}), Q(\vector{X}_{it})\}$,
    \item $Q\big({g_0(\vector{X}_{it})}\big)\approx {g}_1\big({Q(\vector{X}_{it})}\big)$ is learnt iteratively from $\{Q(Y_{it}), Q(\vector{X}_{it})\}$ for model~\eqref{eqn:plr-iv}, relying on the estimates of ${l}_1$ and ${m}_1$.
\end{itemize}
\vspace{.5mm}
With this type of panel approach, the \emph{approximation error} may be large when the nuisance functions are highly nonlinear and non-smooth functions of the covariates.

\textbf{The FD Exact Approach.} The procedure is implemented by selecting the argument \code{approx = "fd-exact"} (default), which internally generates the first-order lags of the covariates by \code{panel_id}. The nuisance function
\begin{itemize}
   \item $Q\big({l_0(\vector{X}_{it})}\big)=l_1(\vector{X}_{it})-l_1(\vector X_{it-1})\equiv l_1(\vector{X}_{it},\vector X_{it-1})$ is learnt directly from transformed data $\{Q(Y_{it}), \vector{X}_{it}, \vector X_{it-1}\}$,
   \item $Q\big({m_0(\vector{X}_{it})}\big)= m_1(\vector{X}_{it})-m_1(\vector X_{it-1})\equiv m_1(\vector{X}_{it},\vector X_{it-1})$ is learnt directly from transformed data $\{Q(D_{it}), \vector{X}_{it}, \vector X_{it-1}\}$,
   \item $Q\big({g_0(\vector{X}_{it})}\big)=g_1(\vector{X}_{it})-g_1(\vector X_{it-1})\equiv g_1(\vector{X}_{it},\vector X_{it-1})$ is learnt iteratively from $\{Q(Y_{it}), \vector{X}_{it}, \vector X_{it-1}\}$ for model~\eqref{eqn:plr-iv}, relying on the estimates of ${l}_1$ and ${m}_1$.
\end{itemize}

\subsection{Score Function}
The second stage of the DML procedure requires the construction of Neyman orthogonal score functions with the predicted nuisance functions to consistently estimate the structural parameter. 


The Neyman orthogonal score function for PLPR models has generic form 
\begin{equation}\label{eqn:sec_score} 
    \psi^\perp(W_i;\theta_0,\bm{\eta}_0)={\vector V}_i^{\perp'}\Sigma_0\inv(\matrix X_i)\mathbf{r}_i , 
\end{equation} 
\noindent where $W_i=\{W_{it}: t=1,\dots,T\}$ with $W_{it} = \{Y_{it},D_{it},\vector X_{it}\}$; ${\vector V}^\perp_i=(V_{i1}^\perp, \dots, V^\perp_{it},\ldots, V^\perp_{iT})'$ contains the orthogonalized regressors chosen to ensure Neyman orthogonality; the row vector $\mathbf{r}_i=(r_{i1},\dots,r_{it},\dots,r_{iT})^\prime$ contains the residuals of the model (either \eqref{eqn:cre_y})-\eqref{eqn:cre_v}) or  \eqref{eqn:Q_y})-\eqref{eqn:Q_d})); $\Sigma_0(\matrix X_i)=E[\mathbf{r}_i\mathbf{r}'_i\mid X_i]$ is the (potentially) heteroskedastic residual variance-covariance matrix; and $\matrix{X}_i=(\vector X_{i1},\dots,\vector X_{it},\dots,\vector X_{iT})$ is the appropriately chosen set of predictor variables. 

The  \code{new()} method allows the user to choose between two types of scores, the  IV-type score and the partial-out (PO) score, which correspond to the untransformed Model~\eqref{eqn:plr-iv} and the untransformed Models~\eqref{eqn:plr_y}–\eqref{eqn:plr_v}, respectively. 

For \code{score = "orth-PO"} (default), Equation~\eqref{eqn:sec_score} takes the following form
\begin{enumerate}[label=(\alph*)]
\item \textbf{CRE:} $r_{it}=a_i+U_{it}=Y_{it}-V_{it} \theta_0 - \widetilde{l}_1(\vector X_{it},\overline{\vector X}_{i})$, $V_{it}=D_{it} - \widetilde{m}_1(\vector X_{it},\overline{\vector X}_i) -c_i$ and $V_{it}^\perp=V_{it}$,  under Model~\eqref{eqn:cre_y}-\eqref{eqn:cre_v},  
\item \textbf{FD-exact and WG-approximation:}   $r_{it}=Q(U_{it})=Q(Y_{it})-Q\big(V_{it}\big)\theta_0 - Q\big(l_1(\vector X_{it}) \big)$ and $V_{it}^{\perp}=Q(V_{it})$, under Model~\eqref{eqn:Q_y}-\eqref{eqn:Q_d},
\end{enumerate}
For \code{score = "orth-IV"}), Equation~\eqref{eqn:sec_score} becomes
\begin{enumerate}[(\alph*')] 
\item \textbf{CRE:} $r_{it}=a_i+U_{it}=Y_{it}-D_{it} \theta_0 - \widetilde{g}_1(\vector X_{it},\overline{\vector X}_{i})$, $V_{it}=D_{it} - \widetilde{m}_1(\vector X_{it},\overline{\vector X}_i) -c_i$ and $V_{it}^\perp=V_{it}$, under Model~\eqref{eqn:cre_iv},  
\item \textbf{FD-exact and WG-approximation:}  $r_{it}=Q(U_{it})=Q(Y_{it})-Q\big(D_{it}\big)\theta_0 - Q\big(g_1(\vector X_{it}) \big)$ and $V_{it}^{\perp}=Q(V_{it})$, under Model~\eqref{eqn:Q_iv}.
\end{enumerate}

The main computational distinction between the two score types is that \code{orth-PO} estimates $l_1$ directly, while \code{orth-IV} targets $g_1$ which is obtained iteratively via $l_1$ and  $m_1$. As a result, \code{orth-IV} is more computationally intensive because it requires an additional machine-learning fit.


\subsection{DML Estimator for Panel Data}
The DML estimator based on the Neyman-orthogonal score~(10) solves the finite-sample analogue of the moment condition $\mathbb{E}[\psi^\perp(W_i; \theta_0, \bm{\eta}_0)]=0$ with respect to $\theta_0$, such that
\begin{equation*}\label{eqn:fmom_cond}
  \frac{1}{N}\sum_{i\in W_i} \psi^\perp(W_i; \theta, \widehat{\bm{\eta}})  = 0,
\end{equation*}
\noindent and has closed-form solution
\begin{equation*}\label{eqn:theta_hat}
    \widehat{\theta} = \Bigg(\frac{1}{N} \sum_{i\in W} \widehat{\vector V}_i^{\perp'}\widehat{\Sigma}_0\inv\widehat{\vector v}_i\Bigg)\inv \frac{1}{N} \sum_{i\in W} \widehat{\vector V}_i^{\perp'}\widehat{\Sigma}_0\inv\widehat{\mathbf{r}}_i, 
\end{equation*}
\noindent where 
\begin{enumerate}[label=(\alph*)]
    \item \textbf{PO score}: $\widehat{\mathbf{r}}_i = \vector Y_{i}- \widehat{\vector l_1(\vector X_{it},\overline{\vector X}_{i})}$ and $\widehat{\vector V}_i^{\perp}=\widehat{\vector v}_i=\widehat{\vector V}_i$ for CRE; $\widehat{\mathbf{r}}_i = Q(\vector Y_{i})- \widehat{Q\big(\vector l_1(\vector X_{it})} \big)$ and  $\widehat{\vector V}_i^{\perp}=\widehat{\vector v}_i=Q(\widehat{\vector V}_i)$ for FD exact and WG approximation approaches;
    \item \textbf{IV-type score}: $\widehat{\mathbf{r}}_i = \vector Y_{i} -\widehat{\vector g_1(\vector X_{it},\overline{\vector X}_{i})}$, $\widehat{\vector V}_i^{\perp}=\widehat{\vector V}_i$, and $ \widehat{\vector v}_i = \vector D_i$ for CRE; and $\widehat{\mathbf{r}}_i = Q(\vector Y_{i})- \widehat{Q\big(g_1(\vector X_{it}) \big)}$ and $\widehat{\vector V}_i^{\perp} = Q(\widehat{\vector V}_i)$, and $\widehat{\vector v}_i=Q(\vector D_i)$  for FD exact and WG approximation approaches.
\end{enumerate}

The DML estimator is $\sqrt{N}$-consistent for the population parameter $\theta_0$ with a normal limiting distribution as in \citet[Theorems~3.1 and~3.2]{chernozhukov2018} and approximate variance
\begin{equation*}
    \sigma^2 \equiv J_0\inv\mathbb{E}\big[\psi^\perp(W_i; \theta, \bm{\eta})\psi^\perp(W_i; \theta, \bm{\eta})'\big] J_0\inv  
\end{equation*}
\noindent where $J_0 =\mathbb{E}({\vector V}_i^{\perp'}{\vector v}_i)$.
By default, the \pkg{xtdml} package computes the cluster-robust statistical inference at the \code{panel_id} level, or at the \code{cluster_cols} if specified otherwise.

\subsection{Block-k-fold Sample Splitting}
The third key component of the DML procedure is sample splitting with cross-fitting to reduce the overfitting bias and restore efficiency. 

The argument \code{draw_sample_splitting}  in the \code{new()} method controls whether sample splitting is conducted. When \code{draw_sample_splitting = TRUE} (default),  the sample $\mathcal{W}$ is divided in $k$ folds  (\code{n_folds = 5} as default) of equal size $N_k\equiv |W_k| = N/K$.\footnote{For example, if the sample size is small, splitting it into many folds further reduces the information available in the estimation sample for estimation. In such cases, the complementary sample may differ substantially in its observed characteristics, effectively forcing the learner to extrapolate} To preserve the longitudinal structure of the data, each cross-sectional unit with its time series is randomly allocated in the same fold.  \pkg{xtdml} uses the specified \code{panel_id} in the \code{xtdml_data_from_data_frame} function to randomly assign subjects into $k$ folds. This sampling strategy is referred as  \emph{block-k-fold} sampling.

For each fold $k$, $W_k$ serves as the main sample for estimating the target parameter $\theta_0$, while its complement $W_k^c$ is used to learn the nuisance functions $\bm{\eta}_0$. When the argument \code{apply_cross_fitting = TRUE} (default)  in the \code{new()} method,  cross-fitting is conducted: the roles of the folds are switched at every iteration so that each fold is used both as an estimation sample and as a complementary sample.

\section{Learners, Hyperparameter Tuning and Data Transformation}\label{sec:learners}
While Neyman orthogonality makes the DML estimator insensitive to minor prediction errors in the nuisance parameters,  finding  optimal configurations of hyperparameters remains an important task to bring the learned nuisance functions as close as possible to their population counterpart and reach state-of-the-art performance in treatment effect estimation \citep{machlanski2023,machlanski2024,bach2024hyper}. 

Hyperparameter tuning proceeds with various trials of different configurations of the hyperparameters to tune, and it is a computationally intensive task. Hyperparameter tuning in \pkg{xtdml} is conducted using the tuning methods provided in the \pkg{mlr3tuning} package. By default, the \code{tune()} method uses grid search \citep{bergstra2012} to compare different configurations of hyperparameter values.  The optimizer (\code{tuner}) randomly searches among a specified number of different values to try per hyperparameter (\texttt{resolution}) and stops the optimization when the specified maximum number of evaluations is reached  (\texttt{terminator}). 
Once the optimal configuration of hyperparameters is found, this is stored and then passed to the DML procedure after calling the \code{fit()} method.

The default tuning procedure in \pkg{xtdml} does not tune on folds (\code{tune_on_folds = FALSE}), and passes all data units into the tuning procedure. Note that the composition of the units in the $k$-th fold differs from the corresponding fold in the DML procedure. When tuning on folds (\code{tune_on_folds = TRUE}), only units in the training sample of fold $k$ are used for tuning and subsequently divided in additional $k$ folds by creating inner training and testing samples. 

\begin{table}[t!]
\centering
\scalebox{.9}{
\begin{threeparttable}	
\begin{tabular}{lccc}  
\hline
&\multicolumn{3}{c}{\code{approach = }}\\
&\code{"cre"}&\code{"wg-approx"}&\code{"fd-exact"}\\	
\hline
\vspace{-3mm}\\
Output (Y)&$Y_{it}$&$Q(Y_{it})$&$Q(Y_{it})$\\
Treatment (D)&$D_{it}$&$Q(D_{it})$&$Q(D_{it})$\\
\vspace{-3mm}\\
\multicolumn{4}{l}{Set of controls passed in the nuisance functions:}\\
\vspace{-3mm}\\
\hspace{1mm}\code{transformX = "no"} & $\{{\vector X}_{it}, \mathbf{\overline{X}}_i, \overline{D}_i\}$ & $Q({\vector X}_{it})$ & $\{{\vector X}_{it}, {\vector X}_{it-1}\}$\\
\hspace{1mm}\code{transformX = "poly"}& $\{\breve{\vector X}_{it}, \breve{\overline{\vector X}}_i, \overline{D}_i\}$ & $Q(\breve{\vector X}_{it})$ & $\{\breve{\vector X}_{it}, \breve{\vector X}_{it-1}\}$\\
\hspace{1mm}\code{transformX = "minmax"}& $\{\dot{\vector X}_{it}, \dot{\overline{\vector X}}_i, \dot{\overline{D}}_i\}$ & $Q(\dot{\vector X}_{it})$ & $\{\dot{\vector X}_{it}, \dot{\vector X}_{it-1}\}$\\
\hline
\end{tabular} 
The table displays how the data $\{Y_{it},D_{it},\vector X_{it}\}$ provided by the user is transformed inside \pkg{xtdml} by panel data approach. The matrix of covariates is $\vector X_{it} = (X_{it,1},\dots,X_{it,p})'$. The individual mean of a variable is calculated as follows $\overline{X}_i= \sum_{i=1}^N X_{it}$. $Q$ is a panel data transformation operator, i.e., $Q(X_{it}) = X_{it} - \sum_{i=1}^N X_{it}$ for the within-group transformation, and $Q(X_{it}) = X_{it} -  X_{it-1}$ for the first-difference transformation. $\breve{X}_{it}$ indicates the polynomial expansion of the the relative variable, such as $\{X_{it,j}X_{it,l}, X_{it,j}^q :j,l \in \{1,\ldots,p\}, q=3\}$.
$\dot{X}$ indicates the transformed variable through the min-max transformation, such that $\dot{X}=(X_{max}-X_{min})/X_{min}$.
\end{threeparttable}
}
\caption{Data transformation inside \pkg{xtdml} by \code{approach} and \code{transformX}.}\label{tab:tranf_data}
\end{table}

In practice, base learners require the data to be pre-processed in specific ways to enhance the learning performance. The \pkg{xtdml} package allows the user to choose among three types of data transformations to apply to the covariates $X$ when initializing the data environment, such as \code{transformX = ("no", "minmax", "poly")}. Selecting the option \code{"no"} (default) leaves the covariates untransformed; this option is recommended for tree-based learners. \code{"minmax"} applies the min-max normalization  to the covariates, i.e., $\dot{X}=(X_{max}-X_{min})/X_{min}$; this option is recommended with neural networks. \code{"poly"} adds polynomials up to order three and interactions between all possible combinations of two and three variables, such as $\{X_{it,j}X_{it,k}, X_{it,j}^q : j,k\in \{1,\ldots,p\}, q=3\}$; this option is recommended for penalized regressors such as Lasso.  Table~\ref{tab:tranf_data} shows how the data $\{Y_{it},D_{it},\vector X_{it}\}$ provided by the user is transformed internally by \pkg{xtdml} across panel data approaches.



\section{Implementation with Simulated Data}\label{sec:examples}
This section outlines the main steps for estimating the structural (causal) parameter using the \pkg{xtdml} package, including:
\begin{enumerate}
\item Installing the package,
\item Loading a sample dataset,
\item Initializing the data environment,
\item Initializing the estimation environment,
\item Conducting hyperparameter tuning and DML estimation,
\item Extraction of stored objects,
\item Display of results in customized tables.
\end{enumerate}

\subsection{Installation of the Package}
The \pkg{xtdml} package is available for installation from CRAN:
\begin{CodeChunk}
\begin{CodeInput}
R> install.packages("xtdml")
R> library(xtdml)
\end{CodeInput}
\end{CodeChunk}

After installation, the documentation of the package and available class methods can be accessed by typing:\footnote{It is not possible to run the examples by typing \code{example("xtdml")} because it is an abstract class intended to define the general structure and interface of the package. Examples can be run for the following classes and functions: \code{make\_plpr\_data} to generate a synthetic data frame for partially linear panel regression (PLPR) models; \code{xtdml\_data\_from\_data\_frame} to initialize the data environment; \code{xtdml\_plr} to run the routine that estimates the PLPR model with DML.}
\begin{CodeChunk}
\begin{CodeInput}
R> help(xtdml)
\end{CodeInput}
\end{CodeChunk}

\subsection{Loading Sample Dataset}
To demonstrate the use of the \pkg{xtdml} package, we use simulated data from the built-in function \code{make_plpr_data()}, based on the following data-generating process
\begin{align*}
Y_{it} &= D_{it}\theta_0 + g_0(X_{it}) + \alpha_i + U_{it}\\
D_{it} &= m_0(X_{it}) + \gamma_i + V_{it},
\end{align*}
\noindent where $U_{it} \sim \mathcal{N}(0,1)$, $V_{it}  \sim \mathcal{N}(0,1)$,
$\alpha_i = \rho A_i + \sqrt{1-\rho^2} B_i $ with $A_i\sim \mathcal{N}(3,3)$, $B_i\sim \mathcal{N}(0,1)$, and $ \gamma_i\sim \mathcal{N}(0,5)$. The covariates are distributed as $X_{it,p} \sim A_i + \mathcal{N}(0, 5)$, where $p$ is the number of covariates.  

The nuisance functions are generated as
\begin{align*}
m_0(X_{it}) &= a_1 [X_{it,1} \times \one(X_{it,1}>0)] + a_2 [X_{it,1} \times X_{it,3}]\\
g_0(X_{it}) & = b_1 [X_{it,1} \times X_{it,3}] + b_2 [X_{it,3} \times \one(X_{it,3}>0)],
\end{align*}
\noindent where $a_1=b_2=0.25$ and $a_2=b_1=0.5$, and $\one(.)$ is the indicator operator.

The function \code{make_plpr_data} accepts several arguments, including: the number of cross-sectional observations (\code{n_obs}),  number of time periods (\code{t_per}), number of covariates (\code{dim_x}) value of the treatment effect (\code{theta}), degree of correlation between the unobserved individual heterogeneity and the observable characteristics (\code{rho}). In this example, we generate a data set with 1,000 cross-sectional units observed over 10 time periods (for a total of 10,000 observations in the sample), and 20 control variables. The true treatment effect is 0.5; the parameter that governs the relationship between the unobserved individual heterogeneity and the covariates is 0.8.
\begin{CodeChunk}
\begin{CodeInput}
R> set.seed(1234)
R> df = make_plpr_data(n_obs = 1000, t_per = 10, 
+                      dim_x = 20, theta = 0.5, rho = 0.8)
\end{CodeInput}
\end{CodeChunk}
A built-in example is available by typing \code{example("make_plpr_data")}, which generates a data frame under different parameter choices.

\subsection{Initialization of the Data Environment}
First, the user is required to initialize the data environment using the function  \\
\code{xtdml_data_from_data_frame()}, which accepts the following arguments: data,  set of raw covariates (\code{x_cols}), exogenous treatment (\code{d_cols}), output / dependent variable (\code{y_col}),  panel identifier (\code{panel_id}),  time identifier (\code{time_id}), and  clustering variable (\code{cluster_cols}) if different from \code{panel_id}, panel data transformation approach (\code{approach}), type of transformation to apply to the set of raw covariates (\code{transformX}).\footnote{Note that panel-related arguments, such as \code{approach} and \code{transformX}, allow users to run the \pkg{xtdml} estimation procedures without performing any \emph{ex-ante} data manipulation or panel transformations.}

In this example, we specify to prepare the data for the CRE estimation approach, and to leave the covariates untransformed (suitable for the tree-based method we specify in the next section).
\begin{CodeChunk}
\begin{CodeInput}
R> x_cols = paste0("X", 1:20)
R> obj_xtdml_data = xtdml_data_from_data_frame(df,
+                        x_cols = x_cols,  
+                        y_col = "y", 
+                        d_cols = "d",
+                        panel_id = "id",
+                        time_id = "time",
+                        approach = "cre", 
+                        transformX = "no")
\end{CodeInput}
\end{CodeChunk}
In this example, the panel identifier and the clustering variable coincide, so specifying \code{cluster\_vars = "id"} is unnecessary.

The output of the data environment can be print by typing:\footnote{A \code{plot()} method is
not implemented for data objects because \code{xtdml\_data\_from\_data\_frame()} provides a ``container'' for storing and checking input data, later used in the DML procedure. Therefore, typing \code{obj\_xtdml\_data\$plot()} would return an error.}
\begin{CodeChunk}
\begin{CodeInput}
R> obj_xtdml_data$print()
\end{CodeInput}
\begin{CodeOutput}
================= xtdml Object ==================
------------------ Data summary ------------------
Outcome variable: y
Treatment variable(s): d
Panel identifier: id
Time identifier: time
Cluster variable(s): id
Covariates: X1, X2, X3, X4, X5, X6, X7, X8, X9, X10, X11, X12, X13, X14, X15, 
X16, X17, X18, X19, X20, m_X1, m_X2, m_X3, m_X4, m_X5, m_X6, m_X7, m_X8, m_X9, 
m_X10, m_X11, m_X12, m_X13, m_X14, m_X15, m_X16, m_X17, m_X18, m_X19, m_X20, m_d
No. Observations: 10000
Panel data approach: cre
Type of transformation for X: no
\end{CodeOutput}
\end{CodeChunk}
Another built-in example is available by typing the code \code{example("xtdml\_data\_from\_data\_frame")} in the \proglang{R} console.

\subsection{Initialization of the Estimation Environment}\label{sec:est_env}
As an illustrative example, we use gradient boosting learner to predict the nuisance functions. We specify: the base learner (\code{xgboost}) for the two nuisance functions, the task to perform (\code{regr} as both the outcome and treatment variables are continuous), and the number of trees to grow (\code{nrounds = 100}).\footnote{The number of boosted trees is set to 100 to reduce the computational time, but it is preferable to be 1000 to enhance estimation accuracy.} 
\begin{CodeChunk}
\begin{CodeInput}
R> learner = lrn("regr.xgboost", nrounds = 100)
R> ml_m = learner$clone()
R> ml_l = learner$clone()
\end{CodeInput}
\end{CodeChunk}

We now initialize the DML estimation environment.
\begin{CodeChunk}
\begin{CodeInput}
R> xtdml_obj = xtdml_plr$new(obj_xtdml_data,  
+                ml_l = ml_l, 
+                ml_m = ml_m,
+                n_folds = 5, 
+                score = "orth-PO",
+                dml_procedure ="dml2"
+                draw_sample_splitting = TRUE,
+                apply_cross_fitting = TRUE)
\end{CodeInput}
\end{CodeChunk}
At this stage, the model has not yet been fitted. Consequently, calling the \code{print()} or \code{summary()} methods will issue the warning message \code{"fit() not yet called"}, as no estimation results are available.

An additional built-in example for the classification and regression tree (\code{rpart}) can be run by typing the code \code{example("xtdml\_plr")} in \proglang{R} console.

\subsection{Conducting Hyperparameter Tuning and DML Estimation}\label{sec:tune_est}
In this section, we proceed by setting up the tuning procedure for a subset of the hyperparameters. Tuning warning messages can be suppressed by typing:
\begin{CodeChunk}
\begin{CodeInput}
R> lgr::get_logger("mlr3")$set_threshold("warn")
R> lgr::get_logger("bbotk")$set_threshold("warn")
\end{CodeInput}
\end{CodeChunk}

The code below defines the hyperparameter search space and tuning settings, all of which can be customized by the user.
\begin{CodeChunk}
\begin{CodeInput}
R> param_grid = list("ml_l" = ps(max_depth = p_int(2, 10),
+                                lambda = p_dbl(0, 2)), 
+                    "ml_m" = ps(max_depth = p_int(2, 10),
+                                lambda = p_dbl(0, 2)))
R> tune_settings = list(n_folds_tune = 5, 
+                  rsmp_tune  = mlr3::rsmp("cv", folds = 5),
+                  terminator = mlr3tuning::trm("evals", n_evals = 5),
+                  tuner      = tnr("grid_search", resolution = 10))
R> xtdml_obj$tune(param_set = param_grid, tune_settings = tune_settings)
\end{CodeInput}
\end{CodeChunk}
The model is then fitted using the selected (fixed and tuned) hyperparameters by calling the \code{fit()} method.
\begin{CodeChunk}
\begin{CodeInput}
R> xtdml_obj$fit()
\end{CodeInput}
\end{CodeChunk}
Calling the \code{print()} method displays the \code{xtdml_obj} object.
\begin{CodeChunk}
\begin{CodeInput}
R> xtdml_obj$print()
\end{CodeInput}
\begin{CodeOutput}
================= xtdml_plr Object ==================

------------------ Data summary ------------------
Outcome variable: y
Treatment variable: d
Covariates: X1, X2, X3, X4, X5, X6, X7, X8, X9, X10, X11, X12, X13, X14, X15, 
X16, X17, X18, X19, X20, m_X1, m_X2, m_X3, m_X4, m_X5, m_X6, m_X7, m_X8, m_X9, 
m_X10, m_X11, m_X12, m_X13, m_X14, m_X15, m_X16, m_X17, m_X18, m_X19, m_X20
Panel identifier: id
Time identifier: time
Cluster variable(s): id
No. Observations: 10000
No. Groups: 1000

------------------ Score & algorithm ------------------
Score function: orth-PO
DML algorithm: dml2
Panel data approach: cre
Type of transformation for X: no

------------------ Machine learner ------------------
Learner of nuisance ml_l: regr.xgboost
RMSE of nuisance ml_l : 4.32708
Learner of nuisance ml_m: regr.xgboost
RMSE of nuisance ml_m : 4.75472
Model RMSE: 13.54673

------------------ Resampling ------------------
No. folds: 5
No. folds per cluster: 5
No. repeated sample splits: 1
Apply cross-fitting: TRUE

------------------ Fit summary ------------------
 Estimates and significance testing of the effect of target variables
  Estimate. Std. Error t value Pr(>|t|)    
d   0.46861    0.03052   15.35   <2e-16 ***
---
Signif. codes:  0 '***' 0.001 '**' 0.01 '*' 0.05 '.' 0.1 ' ' 1
\end{CodeOutput}
\end{CodeChunk}
Alternatively, a concise summary of the estimated model can be displayed by calling the \code{summary()} method.\footnote{Note that the \code{plot()} method is not implemented in \pkg{xtdml}. Calling \code{xtdml\_obj\$plot()} returns an warning message indicating that the plotting method is not supported for \pkg{xtdml} objects.}
\begin{CodeChunk}
\begin{CodeInput}
R> xtdml_boost$summary()
\end{CodeInput}
\begin{CodeOutput}
 Estimates and significance testing of the effect of target variables
  Estimate. Std. Error t value Pr(>|t|)    
d   0.46861    0.03052   15.35   <2e-16 ***
---
Signif. codes:  0 '***' 0.001 '**' 0.01 '*' 0.05 '.' 0.1 ' ' 1
\end{CodeOutput}
\end{CodeChunk}

Additional statistical inference can be conducted by calling the method \code{confint()} which calculates the 95\% confidence intervals. 
\begin{CodeChunk}
\begin{CodeInput}
R> xtdml_obj$confint()
\end{CodeInput}
\begin{CodeOutput}
R>   2.5 %   97.5 %
d .4087802 0.5284328
\end{CodeOutput}
\end{CodeChunk}
The lower and upper bounds of the confidence interval can be extracted separately typting:
\begin{CodeChunk}
\begin{CodeInput}
R> xtdml_obj$confint()[,1]
R> xtdml_obj$confint()[,2]
\end{CodeInput}
\begin{CodeOutput}
0.4087802
0.5284328
\end{CodeOutput}
\end{CodeChunk}

\subsection{Extraction of Stored Objects}
The created object \code{xtdml_obj} stores the results after calling the \code{fit()} method. We can extract various estimates, such as the estimated coefficient of the treatment effect, cluster-robust standard error, p-value, root mean squared errors (RMSEs), selected hyperparameters, panel data metadata.
\begin{CodeChunk}
\begin{CodeInput}
R> xtdml_obj$coef_theta
R> xtdml_obj$se_theta
R> xtdml_obj$pval_theta
\end{CodeInput}
\begin{CodeOutput}
0.4686065
0.03052419
3.435766e-53
\end{CodeOutput}
\end{CodeChunk}

Regarding the quality of learners, we can extract the model RMSE and RMSE of the nuisance functions.
\begin{CodeChunk}
\begin{CodeInput}
R> xtdml_obj$model_rmse
R> as.numeric(xtdml_obj$rmses["ml_l"])
R> as.numeric(xtdml_obj$rmses["ml_m"])
\end{CodeInput}
\begin{CodeOutput}
13.54673
4.327085
4.754724 
\end{CodeOutput}
\end{CodeChunk}

Panel-related metadata, such as the number of observations, number of unique subjects, and number of clusters used in the mode, can be extracted by calling:
\begin{CodeChunk}
\begin{CodeInput}
R> xtdml_obj$get_panel_info()
\end{CodeInput}
\begin{CodeOutput}
$n_obs
[1] 10000

$n_subjects
[1] 1000

$n_groups
[1] 1000
\end{CodeOutput}
\end{CodeChunk}
We can also retrieve the selected (fixed and tuned) hyperparameters for each nuisance function, which are used in the prediction phase of DML procedure, using the dedicated method \code{xtdml_obj\$get\_params()}.\footnote{The field \code{\$params} also provides access to selected hyperparameters of each base learner by the nuisance parameter, but it is not strictly equivalent to the \code{\$get\_params()} method. That is, \code{\$params}  stores the raw internal parameter lists for each learner, reflecting the object’s internal state, whereas \code{\$get\_params()} is a public interface that returns validated and processed parameter values after tuning, performing internal checks and incorporating defaults.} 
A valid nuisance parameter must be specified in the method \code{xtdml\_obj\$get_params()}, otherwise \proglang{R} issues an error message. For \code{score = "orth-PO"}, the valid arguments are \code{"ml_l"} and \code{"ml_m"}, which correspond to the nuisance functions $l_0(X) \equiv \mathbb{E}(Y | X)$ and $m_0(X) \equiv \mathbb{E}(D | X)$, respectively.

The hyperparameters used for predicting \code{"ml_l"} are:
\begin{CodeChunk}
\begin{CodeInput}
R> xtdml_obj$get_params("ml_l")
\end{CodeInput}
\begin{CodeOutput}
$nrounds
[1] 100

$nthread
[1] 1

$verbose
[1] 0

$max_depth
[1] 8

$lambda
[1] 0.2222222
\end{CodeOutput}
\end{CodeChunk}
The hyperparameters used for predicting \code{"ml_m"} are:
\begin{CodeChunk}
\begin{CodeInput}
R> xtdml_obj$get_params("ml_m")
\end{CodeInput}
\begin{CodeOutput}
$nrounds
[1] 100

$nthread
[1] 1

$verbose
[1] 0

$max_depth
[1] 6

$lambda
[1] 0.8888889
\end{CodeOutput}
\end{CodeChunk}

\subsection{Display of Results in Customized Tables}
The stored results can be displayed in a customized table as follows.
\begin{CodeChunk}
\begin{CodeInput}
R> table = matrix(0, 10, 1)
R> table[,1] = cbind(xtdml_obj$coef_theta,
+            xtdml_obj$se_theta,
+            xtdml_obj$pval_theta,
+            ci[ , 1], ci[ , 2],
+            xtdml_obj$model_rmse,
+            as.numeric(xtdml_obj$rmses["ml_l"]),
+            as.numeric(xtdml_obj$rmses["ml_m"]),
+            xtdml_obj$get_panel_info()$n_obs,
+            xtdml_obj$get_panel_info()$n_groups)
R> rownames(table)= c("Estimate", "Std. Error", "P-value", 
+                     "Lower bound 95% CI", "Upper bound 95% CI",
+                     "Model RMSE", "MSE of l", "MSE of m",
+                     "No. observations","No. groups")
R> colnames(table)= c("xtdml-xgboost")
R> formatted_table = matrix("", nrow=10, ncol=1)
R> formatted_table[1:8, 1] = formatC(table[1:8, 1], digits=4, format="f")
R> formatted_table[9:10, 1] = formatC(table[9:10, 1], digits=0, format="f")
R> rownames(formatted_table) = rownames(table)
R> colnames(formatted_table) = colnames(table)
R> print(formatted_table, quote = FALSE)
\end{CodeInput}
\begin{CodeOutput}
                   xtdml-xgboost
Estimate           0.4686       
Std. Error         0.0305       
P-value            0.0000       
Lower bound 95% CI 0.4088       
Upper bound 95% CI 0.5284       
Model RMSE         13.5467      
MSE of l           4.3271       
MSE of m           4.7547       
No. observations   10000        
No. groups         1000    
\end{CodeOutput}
\end{CodeChunk}
Similarly, the analysis can be repeated using the IV-type orthogonal score. This requires setting \code{score = "orth-IV"} when calling \code{xtdml\_plr\$new()}. For brevity, the corresponding code is omitted here but is provided in the replication package. The resulting estimation output is reported below.
\begin{CodeChunk}
\begin{CodeOutput}
                   xtdml-xgboost
Estimate           0.5569       
Std. Error         0.0165       
P-value            0.0000       
Lower bound 95% CI 0.5246       
Upper bound 95% CI 0.5892       
Model RMSE         10.9366      
MSE of l           4.4232       
MSE of m           4.6265       
No. observations   10000        
No. groups         1000   
\end{CodeOutput}
\end{CodeChunk}
\newpage
\section{Empirical Example}\label{sec:empirical}
To illustrate the use of the \pkg{xtdml} package, we use the \code{Produc} panel dataset, available in the \pkg{plm} package in \proglang{R}. The \code{Produc} dataframe contains information on 48 US states over 1970-1986 (17 years) in long format, including:
\begin{itemize}
    \item state: Name of the U.S. state -- (panel identifier)
    \item year: Year of observation -- (time identifier)
    \item region: Region code (1 to 4) -- (clustering variable)
    \item pcap:	Public capital stock per worker -- (treatment)
    \item pc:	Private capital stock per worker
    \item emp:	Employment (in thousands)
    \item unemp:	Unemployment rate
    \item gsp:	Gross State Product
    \item hwy:	Highway capital
    \item water:	Water capital
    \item util:	Utility capital
\end{itemize}
In this empirical example, we compare the effect of increased public investment (\code{pcap}) on state-level economic growth (\code{gsp}) using different base learners and panel data approaches within the DML framework for panel data. The nuisance functions, \code{ml_l} and \code{ml_m},  are learnt  using three base learners from \pkg{mlr3} ecosystem: boosted trees (\code{xgboost}), cross-validated Lasso (\code{cv_glmet}), and neural networks (\code{nnet}). We select the partial-out score function (\code{score = 'orth-PO'}), 3 folds (\code{n_folds = 3}),  with cross fitting (\code{apply_cross_fitting = TRUE}).
This example allows us to compute clustered statistical inference at a level (the region) different from the panel identifier (the US states).

We first load data from \pkg{plm} package.
\begin{CodeChunk}
\begin{CodeInput}
R> library(plm)
R> data("Produc", package = "plm")
\end{CodeInput}
\end{CodeChunk}

We then generate a binary treatment variable by defining treated units as those above the median value of \code{pcap}.
\begin{CodeChunk}
\begin{CodeInput}
R> threshold = median(Produc$pcap, na.rm = TRUE)
R> Produc$treated = ifelse(Produc$pcap > threshold, 1, 0)
\end{CodeInput}
\end{CodeChunk}

Generate the logarithm of the variables and define output variable, treatment variables, and controls to pass into the DML environment.
\begin{CodeChunk}
\begin{CodeInput}
R> Produc = Produc %>%
+          mutate(l_pcap = log(pcap),
+                 l_pc   = log(pc),
+                 l_emp  = log(emp),
+                 l_gsp  = log(gsp),
+                 l_hwy  = log(hwy),
+                 l_water = log(water),
+                 l_util = log(util))         
R> yvar  = "l_gsp"
R> dvar  = "treated"
R> xvars = c("emp", "pc", "hwy", "water", "util")
R> ln_xvars = c("l_emp", "l_pc", "l_hwy", "l_water", "l_util")
\end{CodeInput}
\end{CodeChunk}

In the next sections, we report the code to estimate the structural parameter using the FD exact approach, and the estimation results obtained with the CRE and WG approximation approaches; detailed code is available in the replication package.



\subsection{Example with Gradient Boosting}
We use the \code{xgboost} learner available in \pkg{mlr3extralearners}, and fix the number of boosted trees to 100 to reduce the computational time.
\begin{CodeChunk}
\begin{CodeInput}
R> learner = lrn("regr.xgboost", nrounds = 100)
R> ml_m = learner$clone()
R> ml_l = learner$clone()
\end{CodeInput}
\end{CodeChunk}

We initialize the data environment selecting the options of estimating the model using the FD exact approach and raw variables without any transformation (\code{transformX = "no"}).
\begin{CodeChunk}
\begin{CodeInput}
R> obj_xtdml_data_boost = xtdml_data_from_data_frame(df,
+                              x_cols = c(xvars,pvars),  
+                              y_col = yvar,
+                              d_cols = dvar,
+                              panel_id = "county", 
+                              time_id = "year",
+                              approach = "fd-exact", 
+                              transformX = "no")
\end{CodeInput}
\end{CodeChunk}
We initialize the DML estimation environment using the data object created above.
\begin{CodeChunk}
\begin{CodeInput}
R> xtdml_boost = xtdml_plr$new(obj_xtdml_data_boost,
+                              ml_l = ml_l, 
+                              ml_m = ml_m,
+                              n_folds = 3,
+                              score = "orth-PO")
\end{CodeInput}
\end{CodeChunk}

We define the ranges of the the hyperparameters to tune tune via grid search (\code{lambda} is the L2 regularization term on weights; and \code{max_depth} is the maximum depth of the tree), and the relative tuning options (\code{terminator} is the stop criterion of the tuning process after \code{n\_evals} combinations of values to try,  and \code{resolution} is the number of grid points generated per hyperparameter to tune).
\begin{CodeChunk}
\begin{CodeInput}
R> param_grid = list("ml_l" = ps(max_depth = p_int(2, 10),
+                                lambda = p_dbl(0, 5)),
+                    "ml_m" = ps(max_depth = p_int(2, 10),
+                                lambda = p_dbl(0, 5)))
R> tune_settings = list(n_folds_tune = 3, 
+                       rsmp_tune  = mlr3::rsmp("cv", folds = 3),
+                       terminator = mlr3tuning::trm("evals", n_evals = 10),
+                       tuner      = tnr("grid_search", resolution = 10))
R> xtdml_boost$tune(param_set = param_grid, tune_settings = tune_settings)
\end{CodeInput}
\end{CodeChunk}

We fit the model and print summary of results.
\begin{CodeChunk}
\begin{CodeInput}
R> xtdml_boost$fit()
R> xtdml_boost$summary()
\end{CodeInput}
\begin{CodeOutput}
    Estimates and significance testing of the effect of target variables
      Estimate. Std. Error t value Pr(>|t|)  
 treated -0.004145   0.010417  -0.398    0.691
Signif. codes:  0 ‘***’ 0.001 ‘**’ 0.01 ‘*’ 0.05 ‘.’ 0.1 ‘ ’ 1
\end{CodeOutput}
\end{CodeChunk}

\subsection{Example with Lasso}
As a second learner, we use cross-validated Lasso (\code{cv_glmnet}), where the optimal value of the hyperparameter \code{lambda} is selected as the one that minimizes the cross-validated error.
\begin{CodeChunk}
\begin{CodeInput}
R> learner = lrn("regr.cv_glmnet", s = "lambda.min")
R> ml_m = learner$clone()
R> ml_l = learner$clone()
\end{CodeInput}
\end{CodeChunk}

We then initialize the data and estimation environments. We select the option to include a polynomials and their interactions of the raw covariates (\code{transformX = "poly"}) to satisfy Lasso weak sparsity assumption.
\begin{CodeChunk}
\begin{CodeInput}
R> obj_xtdml_data_lasso = xtdml_data_from_data_frame(df,
+                              x_cols = c(xvars,pvars),  
+                              y_col = yvar,
+                              d_cols = dvar,
+                              panel_id = "county", 
+                              time_id = "year",
+                              approach = "fd-exact", 
+                              transformX = "poly")
R> xtdml_lasso = xtdml_plr$new(obj_xtdml_data_lasso,
+                              ml_l = ml_l, 
+                              ml_m = ml_m,
+                              n_folds = 3) 
\end{CodeInput}
\end{CodeChunk}

We start the DML estimation procedure and print the results.
\begin{CodeChunk}
\begin{CodeInput}
R> xtdml_lasso$fit() 
R> xtdml_lasso$summary()  
\end{CodeInput}
\begin{CodeOutput}
 Estimates and significance testing of the effect of target variables
      Estimate. Std. Error t value Pr(>|t|)  
 treated -0.0003351  0.0124348  -0.027    0.978
Signif. codes:  0 ‘***’ 0.001 ‘**’ 0.01 ‘*’ 0.05 ‘.’ 0.1 ‘ ’ 1
   
\end{CodeOutput}
\end{CodeChunk}

\subsection{Example with Neural Networks}
The third learner we employ is a single-layer neural network with 100 maximum number of iterations (\code{maxit}), 1000 maximum allowable number of weights (\code{MaxNWts}), and no switch for tracing optimization (\code{trace = FALSE}).
\begin{CodeChunk}
\begin{CodeInput}
R> ml_l = lrn("regr.nnet", maxit = 100, MaxNWts = 1000, trace = FALSE)
R> ml_m = lrn("regr.nnet", maxit = 100, MaxNWts = 1000, trace = FALSE)
\end{CodeInput}
\end{CodeChunk}

We initialize the data environment and estimation environment. We apply a min-max transformation (\code{transformX = "minmax"}) to ensure that all raw covariates are on a comparable scale. This improves the stability and convergence of neural networks, which are sensitive to the relative magnitudes of the features.
\begin{CodeChunk}
\begin{CodeInput}
R> obj_xtdml_data_nnet = xtdml_data_from_data_frame(df,
+                             x_cols = c(xvars,pvars),  
+                             y_col = yvar,
+                             d_cols = dvar,
+                             panel_id = "county", 
+                             time_id = "year",
+                             approach = "fd-exact", 
+                             transformX = "minmax")
R> xtdml_nnet = xtdml_plr$new(obj_xtdml_data,
+                             ml_l = ml_l, 
+                             ml_m = ml_m,
+                             n_folds = 3)
\end{CodeInput}
\end{CodeChunk}

We set the hyperparameters to tune: \code{size} being the number of units in the hidden layer, and \code{decay} being the parameter for weight decay. We then initialize the tuning task via grid search.
\begin{CodeChunk}
\begin{CodeInput}
R> param_grid = list("ml_l" = ps(size = p_int(2, 10),
+                           decay = p_dbl(0, 0.05)),
+                    "ml_m" = ps(size = p_int(2, 10),
+                           decay = p_dbl(0, 0.05)))
R> tune_settings = list(n_folds_tune = 3, 
+                       rsmp_tune  = mlr3::rsmp("cv", folds = 3),
+                       terminator = mlr3tuning::trm("evals", n_evals = 10),
+                       tuner      = tnr("grid_search", resolution = 10))
R> xtdml_nnet$tune(param_set = param_grid, tune_settings = tune_settings)
\end{CodeInput}
\end{CodeChunk}

We now fit the DML model using tuned hyperparameters. We finally print the summary of the results.
\begin{CodeChunk}
\begin{CodeInput}
R> xtdml_nnet$fit() 
R> xtdml_nnet$summary()
\end{CodeInput}
\begin{CodeOutput}
    Estimates and significance testing of the effect of target variables
      Estimate. Std. Error t value Pr(>|t|)    
 treated -0.004316   0.011970  -0.361    0.718

Signif. codes:  0 ‘***’ 0.001 ‘**’ 0.01 ‘*’ 0.05 ‘.’ 0.1 ‘ ’ 1
\end{CodeOutput}
\end{CodeChunk}

\subsection{Extraction of DML Objects}
We extract the DML objects relevant for statistical analysis and display them in a customized table.
\begin{CodeChunk}
\begin{CodeInput}
R> table = matrix(0, 9, 3)
R> table[,1] = cbind(xtdml_lasso$coef_theta, 
+        xtdml_lasso$se_theta, 
+        xtdml_lasso$pval_theta, 
+        xtdml_lasso$model_rmse, 
+        as.numeric(xtdml_lasso$rmses["ml_l"]), 
+        as.numeric(xtdml_lasso$rmses["ml_m"]),
+        xtdml_boost$get_panel_info()$n_obs,
+        xtdml_boost$get_panel_info()$n_subjects,
+        xtdml_boost$get_panel_info()$n_groups)
R> table[,2] = cbind(xtdml_boost$coef_theta,
+        xtdml_boost$se_theta,
+        xtdml_boost$pval_theta,
+        xtdml_boost$model_rmse,
+        as.numeric(xtdml_boost$rmses["ml_l"]),
+        as.numeric(xtdml_boost$rmses["ml_m"]),
+        xtdml_lasso$get_panel_info()$n_obs,
+        xtdml_lasso$get_panel_info()$n_subjects,
+        xtdml_lasso$get_panel_info()$n_groups)
R> table[,3] = cbind(xtdml_nnet$coef_theta,
+        xtdml_nnet$se_theta,
+        xtdml_nnet$pval_theta,
+        xtdml_nnet$model_rmse,
+        as.numeric(xtdml_nnet$rmses["ml_l"]),
+        as.numeric(xtdml_nnet$rmses["ml_m"]),
+        xtdml_nnet$get_panel_info()$n_obs,
+        xtdml_nnet$get_panel_info()$n_subjects,
+        xtdml_nnet$get_panel_info()$n_groups)
R> rownames(table)= c("Estimate", "Std. Error", "P-value", 
+        "Model RMSE", "MSE of l", "MSE of m", 
+        "No. observations","No. subjects","No. groups")
R> colnames(table)= c("xtdml-Boosting", "xtdml-Lasso", "xtdml-Nnet")
R> print(table, 3)
\end{CodeInput}
\begin{CodeOutput}
                 xtdml-Boosting xtdml-Lasso xtdml-Nnet
Estimate                 -0.004       0.000     -0.004
Std. Error                0.010       0.012      0.012
P-value                   0.691       0.978      0.718
Model RMSE                0.328       0.396      0.361
MSE of l                  0.053       0.045      0.039
MSE of m                  0.131       0.102      0.102
No. observations        768.000     768.000    768.000
No. subjects             48.000      48.000     48.000
No. groups                9.000       9.000      9.000
\end{CodeOutput}
\end{CodeChunk}

The analysis is also repeated using alternative panel data approaches available in \pkg{xtdml}. For brevity, we omit the code needed to reproduce these additional results; the additional code is provided in the replication package.\footnote{The omitted code loops over the three panel approaches in the following order \code{approaches = c("cre", "wg-approx")}. Note that the classification task (\code{classif}) is possible only for the CRE approach, because the binary treatment becomes discrete after the WG and FD transformations. The analysis in this section can also be repeated choosing the Neyman-orthogonal IV-type score (\code{score = "orth-IV"}), or other base learners or ensembles from \pkg{mlr3}.}

The output for \code{approach = "cre"} is
\begin{CodeChunk}
% \begin{CodeInput}
% \end{CodeInput}
\begin{CodeOutput}
                 xtdml-Boosting xtdml-Lasso xtdml-Nnet
Estimate                  0.120      -0.011     -0.126
Std. Error                0.018       0.549      0.119
P-value                   0.000       0.984      0.290
Model RMSE                1.317      14.312      1.390
MSE of l                  0.167       1.111      0.183
MSE of m                  0.213       0.288      0.154
No. observations        816.000     816.000    816.000
No. subjects             48.000      48.000     48.000
No. groups                9.000       9.000      9.000
\end{CodeOutput}
\end{CodeChunk}

The output for \code{approach = "wg-approx"} is
\begin{CodeChunk}
% \begin{CodeInput}
% \end{CodeInput}
\begin{CodeOutput}
                 xtdml-Boosting xtdml-Lasso xtdml-Nnet
Estimate                  0.059       0.037      0.173
Std. Error                0.024       0.016      0.072
P-value                   0.016       0.018      0.017
Model RMSE                0.681       0.919      1.028
MSE of l                  0.091       0.173      0.143
MSE of m                  0.193       0.356      0.162
No. observations        816.000     816.000    816.000
No. subjects             48.000      48.000     48.000
No. groups                9.000       9.000      9.000
\end{CodeOutput}
\end{CodeChunk}
The overall running time for the code with the FD exact approach is 38.8 seconds, 36.79 seconds for CRE, and 35.44 seconds for WG approximation. 

\subsection{Interpretation of Results}
The DML results show notable variation in estimated effects across learners and panel data transformations. This reflects both the different nonlinear patterns that learners can accommodate and the modelling assumptions behind each panel data transformation. 

The FD exact estimates are more stable across learners than CRE estimates, and noticeably smaller than those obtained with the WG approximation approach. 
For the CRE approximation approaches, the DML estimates obtained with gradient boosting have opposite signs than those produced by Lasso and neural network. This may reflect sensitivity of the learner to the small cross-sectional size ($N = 48$) and the limited number of boosting rounds ($B = 100$). In addition,  the DML point estimates under FD exact and CRE are insignificant for Lasso and neural networks, unlike gradient boosting. Under the WG approximation, the effects are all statistically significant at 5\% level. 

Regarding the predictive performance, model RMSEs from the FD exact estimations are lower and more consistent across learners, indicating greater precision. The RMSEs of the nuisance functions $l$ and $m$ are also smaller for FD exact than for CRE and WG approximation, suggesting that learners more accurately capture the relationships involving $(X_{it}, X_{it-1})$ for FD exact than those based on $(X_{it}, \overline{X}_i)$ for CRE or $Q(X_{it})$ for WG approximation. Finally, neural networks consistently report the lowest RMSEs for  $l$ and $m$, indicating its superior predictive accuracy than the other selected learners, particularly relative to gradient boosting, across all panel-data approaches.

Overall, the FD exact approach is generally more robust, requires no structural assumptions on individual effects, and produces more accurate predictions of the nuisance functions since the learners jointly predict the relationship between the covariates and their lags, which is a more demanding learning task than in CRE or WG settings.

\section{Conclusion} \label{sec:summary}
This article introduces the R package \pkg{xtdml}, which currently implements partially linear panel regression models with high-dimensional confounding variables, low-dimensional fixed-effects and exogenous treatment within the double machine learning framework, as proposed by \citet{clarke2025}. The article documents the functionalities of the \pkg{xtdml} package with a simulated example, and provides practical guidance for its use with real data when different panel data approaches and base learners are chosen. 


\section*{Computational Details}
The results in this paper were obtained using R version 4.5.1 (2025-06-13 ucrt) on a laptop equipped with a 12th Gen Intel Core i7-1260P processor with 12 cores and 16 threads (base clock speed 2.1 GHz), 16 GB of RAM (15.9 GB installed), and running a 64-bit Windows operating system on an x64-based processor. 
The following packages have been used to run the code: \pkg{xtdml} 0.1.11, \pkg{xgboost} 1.7.11.1,  \pkg{nnet} 7.3-20, \pkg{glmnet} 4.1-10,  \pkg{mlr3} 1.3.0,  \pkg{mlr3misc} 0.19.0,   \pkg{mlr3tuning} 1.5.0, \pkg{mlr3learners} 0.13.0, \pkg{mlr3measures} 1.2.0, \pkg{paradox} 1.0.1, \pkg{bbotk} 1.8.1, \pkg{R6} 2.6.1, \pkg{checkmate} 2.3.3, \pkg{clusterGeneration} 1.3.8, \pkg{MASS} 7.3-65, \pkg{MLmetrics} 1.1.3, \pkg{Rcpp} 1.1.0.


\section*{Acknowledgments}
The author is grateful to Modesto Escobar Mercado for his assistance with the submission of the \pkg{xtdml} package to CRAN during his visit to the Institute for Social and Economic Research, and thanks Paul Clarke and Omar Hussein for their valuable comments.
The author also acknowledges that this research was supported by the UK Economic and Social Research Council (ESRC) under award ES/S012486/1 (MiSoC).


\bibliography{refs.bib}


\newpage




\end{document}